\documentclass[journal,twoside,web]{ieeecolor}
\usepackage{tmi}
\usepackage[noadjust]{cite}
\usepackage{amsmath,amssymb,amsfonts}
\usepackage{algorithmic}
\usepackage[pdftex]{graphicx}
\usepackage{textcomp}
\usepackage{lipsum}
\usepackage{mathtools}
\usepackage{cuted}
\usepackage{bm}
\usepackage{array, multirow}
\usepackage{diagbox}

\DeclareMathOperator*{\argmax}{argmax}

\newcolumntype{P}[1]{>{\centering\arraybackslash}p{#1}}
\makeatletter
\def\thickhline{%
  \noalign{\ifnum0=`}\fi\hrule \@height \thickarrayrulewidth \futurelet
   \reserved@a\@xthickhline}
\def\@xthickhline{\ifx\reserved@a\thickhline
               \vskip\doublerulesep
               \vskip-\thickarrayrulewidth
             \fi
      \ifnum0=`{\fi}}
\makeatother

\newlength{\thickarrayrulewidth}
\def\BibTeX{{\rm B\kern-.05em{\sc i\kern-.025em b}\kern-.08em
    T\kern-.1667em\lower.7ex\hbox{E}\kern-.125emX}}
\begin{document}
\title{Efficient Pairwise Neuroimage Analysis using the Soft Jaccard Index and 3D Keypoint Sets}
\author{Laurent~Chauvin, Kuldeep~Kumar, Christian~Desrosiers, William~Wells~III and Matthew~Toews
\thanks{Manuscript submitted March 15, 2021. This work was supported by NIH grant P41EB015902 (NAC), the Quebec (FRQNT) New Researchers Startup Program and the Canadian National Sciences and Research Council (NSERC) Discovery Grant. Data were provided by the Human Connectome Project, WU-Minn Consortium (Principal Investigators: David Van Essen and Kamil Ugurbil; 1U54MH091657) funded by the 16 NIH Institutes and Centers that support the NIH Blueprint for Neuroscience Research; and by the McDonnell Center for Systems Neuroscience at Washington University.}

\thanks{L. Chauvin, C. Desrosiers and M. Toews are with the Ecole de Technologie Superieure, Montreal, QC H3C 1K3, Canada (email: laurent.chauvin.2@etsmtl.net, christian.desrosiers@etsmtl.ca, matt.toews@gmail.com).}
\thanks{K. Kumar is with CHU Sainte-Justine Research Center, University of Montreal, Montreal, QC H3T 1C5, Canada (email: kuldeep.kumar@umontreal.ca).}
\thanks{W. Wells is with the Harvard Medical School, Boston, MA 02115, USA and the Massachusetts Institute of Technology, Boston, MA 02139, USA (email: sw@bwh.harvard.edu).}}

\maketitle

\begin{abstract}
We propose a novel pairwise distance measure between image keypoint sets, for the purpose of large-scale medical image indexing. Our measure generalizes the Jaccard index to account for soft set equivalence (SSE) between keypoint elements, via an adaptive kernel framework modeling uncertainty in keypoint appearance and geometry. A new kernel is proposed to quantify the variability of keypoint geometry in location and scale. Our distance measure may be estimated between $O(N^2)$ image pairs in $O(N~\log~N)$ operations via keypoint indexing. Experiments report the first results for the task of predicting family relationships from medical images, using 1010 T1-weighted MRI brain volumes of 434 families including monozygotic and dizygotic twins, siblings and half-siblings sharing 100\%-25\% of their polymorphic genes. Soft set equivalence and the keypoint geometry kernel improve upon standard hard set equivalence (HSE) and appearance kernels alone in predicting family relationships. Monozygotic twin identification is near 100\%, and three subjects with uncertain genotyping are automatically paired with their self-reported families, the first reported practical application of image-based family identification. Our distance measure can also be used to predict group categories, sex is predicted with an AUC=0.97. Software is provided for efficient fine-grained curation of large, generic image datasets.
\end{abstract}


\section{Introduction}
\label{sec:introduction}
\IEEEPARstart{H}{ealth} treatment is increasingly personalized, where treatment decisions are conditioned on patient-specific information in addition to knowledge regarding the general population~\cite{hamburg2010path}. Modern genetic testing allows us, based on large libraries of human DNA samples, to cheaply predict patient-specific characteristics, including immediate family relationships, and also characteristics shared across the population including racial ancestry, sex, hereditary disease status, etc.~\cite{annas201423andme,10002015global}. 
The brain is the center of cognition and a complex organ, tightly coupled to the genetic evolution of animals and in particular that of the human species. To what degree is the human brain phenotype coupled to the underlying genotype? How does the brain image manifold vary locally with genotype, i.e. immediate relatives sharing 25-100\% of their polymorphic genes, or between broad groups defined by subtle genetic factors such as racial ancestry or sex?

Large, publicly available databases allow us to investigate these questions from aggregated MRI and genetic data, e.g. 1000+ subjects~\cite{VanEssen2012}. The shape of the brain has been modeled as lying on a smooth manifold in high dimensional MRI data space~\cite{Gerber2010ManifoldAnalysis.,Brosch2013ManifoldMRIs}, where phenotype can be described as a smooth deformation conditioned on developmental factors including the environment. However the brain is also naturally described as a rich collection of spatially localized neuroanatomical structures, including common structures such as the basal ganglia shared across the population, but also highly specific patterns such as cortical folds that may only be observable in specific individuals or close family members~\cite{ono1990atlas}.

\begin{figure}[!h]
  \centering
   \includegraphics[width=1\linewidth,draft=false]{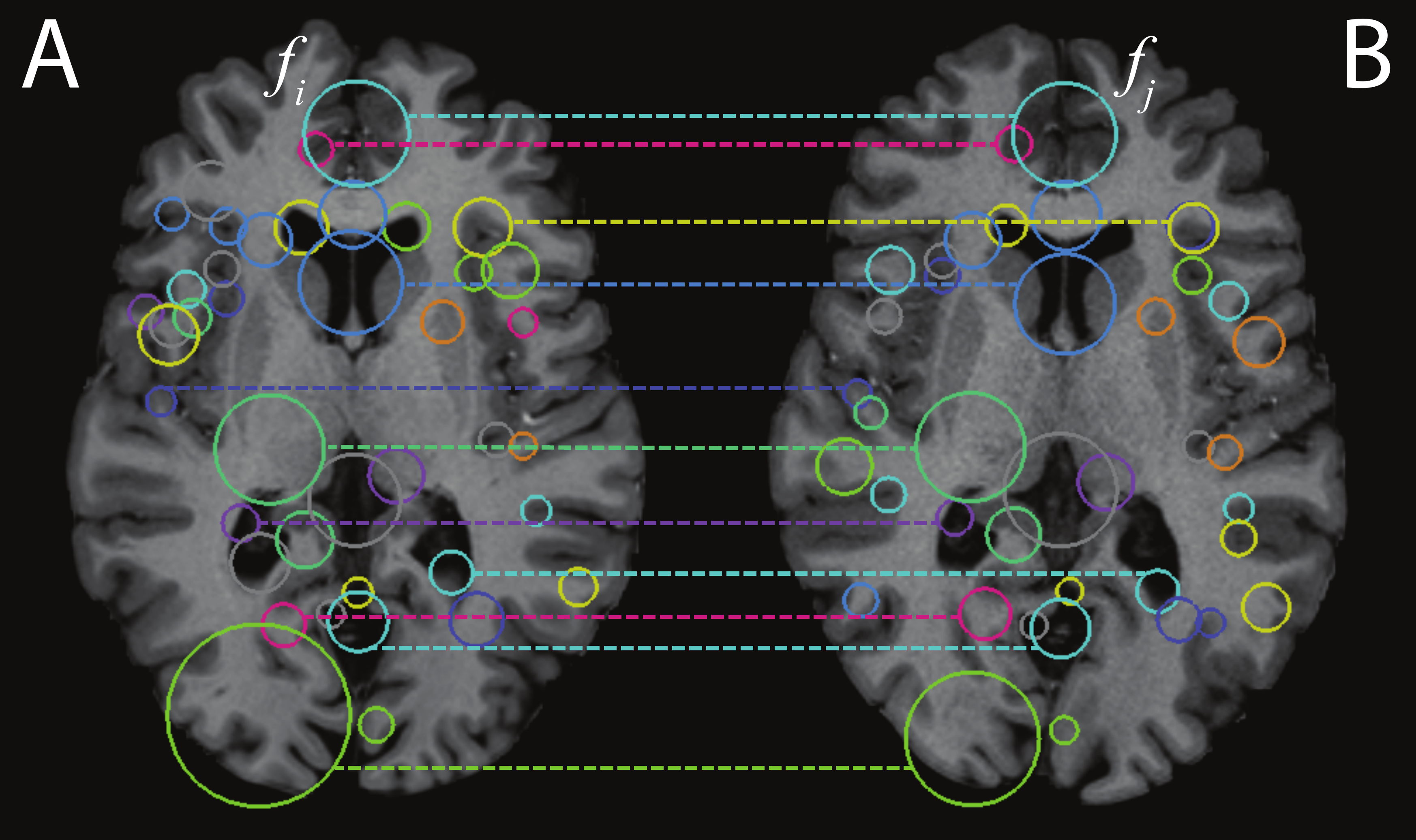}
   \caption{Illustrating two sets of 3D keypoints $A=\{f_i\}$ and $B=\{f_j\}$ extracted from MRI brain volumes of monozygotic twins. Colored circles represent the 3D locations $\bar{x} \in R^3$ and scales (radii) $\sigma \in R^1$ of keypoints in 2D axial slices, and links are correspondences identified via the SIFT-Rank~\cite{Toews2013a} algorithm, note the degree of similarity between keypoint pairs. }
  \label{fig:sift_visualization}
\end{figure}

The keypoint representation is an intuitively appealing means of modeling specific, localized phenomena, i.e. a set of descriptors automatically identified at salient image locations as shown in Figure~\ref{fig:sift_visualization}. A keypoint set can be viewed as an element of a high-dimensional manifold, and medical imaging  applications such as regression or classification can be formulated in terms of a suitable geodesic distance between sets. As keypoint sets are variable sized, typical metrics based on fixed-length vectors such as L-norms~\cite{Gerber2010ManifoldAnalysis.,Brosch2013ManifoldMRIs} do not readily apply. Distances defined based on set overlap measures as the Jaccard index~\cite{Levandowsky1971DistanceSets} have proven to be effective in recent studies investigating genetics and brain MRI~\cite{Toews2016,Chauvin2020NeuroimageRelatives}. For example, by defining set equivalence in terms of nearest neighbor (NN) keypoint correspondences, the Jaccard distance may be used to predict pairwise relationships. Nevertheless, the assumption of binary or hard set equivalence (HSE) between keypoints is a crude approximation given probabilistic uncertainty inherent to natural image structure, and is ill-defined for variable sized datasets where the number of NN correspondences may be variable or unknown a-priori.

Our primary contribution is a new pairwise distance measure for 3D keypoint sets, generalizing the Jaccard index to account for soft set equivalence (SSE) between keypoint elements. SSE is achieved via a kernel framework for keypoint similarity, and we introduce a new kernel that normalizes pairwise keypoint displacement by keypoint scale, accounting for localization uncertainty in scale-space. Our work extends and compares against keypoint-based neuroimaging analysis methods~\cite{Toews2016,Kumar2018Multi-modalFramework,Chauvin2020NeuroimageRelatives}, specifically the keypoint signatures approach~\cite{Chauvin2020NeuroimageRelatives}, currently the only method reporting perfect accuracy at identifying scans of the same individual in large public neuroimage MRI datasets, e.g. OASIS~\cite{marcus2007open}, ADNI~\cite{jack2008alzheimer} and HCP~\cite{van2013wu}. Here, following preliminary work~\cite{Chauvin2019AnalyzingManifold}, we demonstrate that the soft Jaccard distance is specific enough to predict family relationships from medical images, and we present the first results for this new task using 1010 MRIs of 434 families from the HCP~\cite{van2013wu} dataset. The entire method is based on highly efficient keypoint indexing, and scales to large datasets of volumetric image data.

% and instances of previously unknown subject labeling consistencies in widely used, public neuroimage MRI datasets (e.g. OASIS~\cite{marcus2007open}, ADNI~\cite{jack2008alzheimer}, HCP~\cite{van2013wu}). 

%This does not belong in the intro: It should be noted that the visualization of the results differ between~\cite{Chauvin2019AnalyzingManifold} and this work, for a better visualuation of outliers.

\section{Related Work}

Our work is motivated both by the study of the link between genotype and the human brain phenotype from large datasets~\cite{sabuncu2016morphometricity}, and by practical applications such as maintaining accurate patient records in hospital Picture Archiving and Communication Systems (PACS).
We adopted a memory-based learning model where all data are stored in memory and accessed via highly-specific keypoint queries. Memory-based learning requires no explicit training procedure~\cite{boiman2008defense}, and approaches Bayes optimal performance as the number of data $N$ becomes large~\cite{Cover1967}. 3D SIFT keypoints~\cite{Toews2013a} derived from Gaussian scale-space theory~\cite{Lowe2004DistinctiveKeypoints,Lindeberg1998} are invariant to global similarity transforms and contrast variations, and thus account for misalignment and scanner variability between images. Efficient indexing reduces the quadratic $O(N^2)$ complexity of nearest neighbor (NN) keypoint lookup to $O(N~\log~N)$, e.g. via $O(\log N)$ KD-tree indexing~\cite{Muja2014}, ensuring that the method scales gracefully to large datasets, e.g. 7500 brain MRIs~\cite{Chauvin2020NeuroimageRelatives} or 20000 lung CTs~\cite{Toews2015}.

Our method seeks a pairwise distance measure between 3D keypoint sets. As different images generally contain different numbers of elements, a natural choice for pairwise distance between variable sized sets are measures based on set intersection or overlap, e.g. the Jaccard distance metric~\cite{Levandowsky1971DistanceSets} first proposed in~\cite{marczewski1958certain} or the Tanimoto distance~\cite{Bajusz2015WhyCalculations,rogers1960computer}. In medical image analysis, the Jaccard distance is typically used to assess pixel-level segmentation accuracy~\cite{Yuan2017AutomaticDistance}, and has proven highly effective for assessing pairwise neuroimage similarity from 3D SIFT keypoint data~\cite{Toews2016,Kumar2018Multi-modalFramework}, where hard set equivalence is determined by NN keypoint descriptor correspondences.
Set theory is predicated on binary equivalence between set elements, which is difficult to justify in the case of keypoints extracted from image data. Soft set theory was been developed to account for non-binary equivalence~\cite{molodtsov1999soft,park2012some,Gardner2014}, where set operations including intersection and union are defined in terms of non-binary equivalence ranging from [0,1]. Jaccard-like distance was used for retrieving near-duplicate 2D photos~\cite{Chum2008NearWeighting}, where the ratio of soft intersection and union operators was used to account for  document frequency. Our work here extends the Jaccard distance proposed in~\cite{Chauvin2020NeuroimageRelatives} to include keypoint geometry in 3D Euclidean metric space coordinates, and compares to the HSE Jaccard measure used in~\cite{Toews2016,Kumar2018Multi-modalFramework}, showing a significant improvement in predicting family relationships.

% Why use SIFT / Local descriptors
The keypoint methodology has a number of advantages, such as robustness to occlusion or missing correspondences, invariance to translation, scaling, rotation and intensity contrast variations between images. Our work adopts the 3D SIFT-Rank~\cite{Toews2013a} representation, a robust, general tool used for a variety of imaging tasks, including registration~\cite{Toews2013a,Machado2018Non-rigidMatching}, whole body segmentation~\cite{wachinger2018keypoint}, classification~\cite{Toews2015,Toews2010}, regression~\cite{toews2012feature}, without the need for context-specific training procedures or data. As in the original 2D SIFT approach, keypoint geometry is represented as a location $\bar{x}$ and scale $\sigma$, these are detected as extrema of a difference-of-Gaussian scale-space~\cite{Lowe2004DistinctiveKeypoints}, as $\argmax_{\bar{x},\sigma} \ |I(\bar{x},\sigma) \, - \, I(\bar{x},\kappa\sigma)|$, approximating the Laplacian-of-Gaussian~\cite{Lindeberg1998},  where $I(\bar{x},\sigma)=I(\bar{x})*\mathcal{N}(\bar{x},\sigma^2)$ represents the image convolved with the Gaussian kernel $\mathcal{N}(\bar{x},\sigma^2)$. The geometry of a keypoint may thus be characterised as an isotropic Gaussian density centered on $\bar{x}$ with variance $\sigma^2$ representing spatial extent $\sigma$ in 3D Euclidean space. Rotationally symmetric Gaussian filtering and uniformly sampled derivative operators lead to scale and rotation invariance. The 3D SIFT-Rank descriptor is a 64-element histogram of rank-ordered oriented gradients (HOG)~\cite{Toews2013a}, sampled about a scale-normalized reference frame centered on $\bar{x}$ and quantized uniformly into 2x2x2=8 spatial locations x 8 orientations. Rank transformation provides invariance to arbitrary monotonic variations of descriptor element values~\cite{Toews2009}.

The 3D SIFT-Rank keypoint method we use is a robust and well-established baseline, however our distance measure may generally be used with any keypoint representation. The search for new keypoint detectors and descriptors remains an active research focus, primarily in the context of 2D computer vision. We refer the reader to an extensive literature review of traditional hand-crafted solutions~\cite{Mukherjee2015ADescriptors} and a recent discussion of 2D SIFT matching technology~\cite{bellavia2020there}. Deep learning keypoint methods have been investigated, examples include LIFT~\cite{Yi2016LIFT:Transform}, DISK~\cite{Tyszkiewicz2020DISK:Gradient}, LF-Net~\cite{Ono2018LF-Net:Images}, SuperPoint~\cite{Detone2018SuperPoint:Description}. Nevertheless, variants of the original SIFT histogram descriptor, including SIFT-Rank~\cite{Toews2009}, DSP-SIFT~\cite{Dong2015} or RootSIFT~\cite{Arandjelovic2012ThreeRetrieval} remain competitive with descriptors derived from training~\cite{Balntas2017HPatches:Descriptors, Schonberger2017ComparativeFeatures}, particularly for non-planar objects~\cite{bellavia2020there}, while requiring no training and few hyperparameters. Learning-based approaches often rely on hand-crafted keypoint detectors to generate training data~\cite{Yi2016LIFT:Transform,Detone2018SuperPoint:Description}. The challenges of developing deep keypoint architectures include training of individual components (i.e. keypoint detector, orientation estimator, descriptor), achieving invariance to rotation or scale changes, the need to retrain for different imaging modalities and body parts, and avoiding bias due to training data procedures~\cite{Geirhos2019Imagenet-trainedRobustness}.  

% Neuro fingerprinting

In terms of neuroimage analysis, brain fingerprinting methods have been used to investigate the variability of individuals based on neuroimage-specific preprocessing and data, e.g. brain segmentations from structural MRI~\cite{Wachinger2015BrainPrint:Morphology,valizadeh2018identification}, neuroanatomic parcellations and functional MRI~\cite{Finn2015}, fiber tracts from diffusion MRI~\cite{kumar2017fiberprint}. None of these methods have been sufficiently specific to achieve perfect accuracy at individual identification, and none have been used for the more difficult task of predicting family members. Heritability studies have investigated correlations in MRIs of siblings, including structural~\cite{thompson2001genetic,vanderLee2017GrayMorphometry} and functional MRI~\cite{Colclough2017}, however these are not designed for prediction. In contrast, the keypoint methodology we adopt may be applied to generic imaging data with no preprocessing, representing distinctive patterns potentially present only in images of specific individuals or immediate family members. Our work here extends and compares against the keypoint-based neuroimage signatures method~\cite{Chauvin2020NeuroimageRelatives}, the only method reporting perfect accuracy at same-subject image prediction from standard neuroimage datasets (e.g. OASIS~\cite{marcus2007open}, ADNI~\cite{jack2008alzheimer}, HCP~\cite{van2013wu}).

Our results here are the first in the literature for the task of family member prediction from volumetric medical image data. Family member prediction has been investigated in the context of 2D face photographs, based on clustering~\cite{kiley2020my}, conditional random fields~\cite{dai2015family}, deep neural networks~\cite{ahmad2019deep}, discriminant analysis~\cite{Juefei-Xu2013AnFeatures}. These solutions are not directly related to our work, as they generally address aspects and challenges specific to the context of 2D photography, e.g. facial expression and lighting conditions~\cite{Paone2014DoubleRecognition}, multiple images per person, group photos of relatively few families (e.g. 16)~\cite{dai2015family}. Accuracy is not generally available for siblings based on genotyping, and near 100\% accuracy for monozygotic twin identification from face photographs has not been reported.

\section{Method}

We seek a pairwise distance measure between two sets $A = \{f_i\}$ and $B = \{ f_j\}$ that can be used to estimate proximity and thus genetic relationships between subjects from image data. We first describe our method in terms of general set theory, then later include details pertaining to set elements $f_i$ defined as 3D image keypoints. Our measure begins with the Jaccard index or intersection-over-union $J_{HSE}(A,B)$ defined as
\begin{align}
    J_{HSE}(A,B) &= \frac{|A \cap B|}{|A| + |B| - |A \cap B|},
    \label{eq:jaccard}
\end{align}
where in Equation~\eqref{eq:jaccard}, $J_{HSE}(A,B)$ is defined by binary or hard set equivalence (HSE) between set elements. $A \cap B$ is the intersection operator between sets $A$ and $B$, and $|A \cap B|$, $|A|=|A \cap A|$, $|B|=|B \cap B|$ represent set cardinality operators that count the numbers of elements present in each set, where $|A \cap B| \le \min\{ |A|, |B|\}$. The Jaccard index is a similarity measure, and may be used to define various distance measures including the Jaccard distance metric~\cite{Levandowsky1971DistanceSets} $1-J_{HSE}(A,B)$ or the Tanimoto distance $-\log J_{HSE}(A,B)$. 

In the case of sets of real data samples, for example keypoint descriptors, binary equivalence may be difficult to establish due to noise or uncertainty in the measurement process, and we seek to redefine $J_{HSE}(A,B)$ in Equation~\eqref{eq:jaccard} such that it more accurately accounts for non-binary equivalence between set elements. This may generally be accomplished by redefining the standard set intersection cardinality operator $|A \cap B|$ in Equation~\eqref{eq:jaccard} by an analogous soft cardinality operator $\mu(A \cap B)$, leading to the following expression for our proposed Jaccard index based on soft set equivalence $J_{SSE}(A,B)$
\begin{equation}
	J_{SSE}(A,B) = \frac{\mu(A \cap B)}{\mu(A) + \mu(B) - \mu(A \cap B)} \label{eq:soft_jaccard}.
\end{equation}
In Equation~\eqref{eq:soft_jaccard}, $J_{SSE}(A,B)$ is defined by the soft cardinality operator $\mu(A \cap B)$ as described in the following section, including cardinality operators $\mu(A)=\mu(A \cap A)$, $\mu(B)=\mu(B\cap B)$. In order to ensure that $J_{SSE}$ remains bounded to the range $[0,1]$, it is important that $\mu(A \cap B)$ be upper bounded by the minimum of the individual soft set cardinalities $\mu(A \cap B) \le \min \left\{ \mu(A), \mu(B)\right\}$.  \\

\noindent {\bf Defining $\boldsymbol{\mu(A \cap B)}$:}
The cardinality of the soft set intersection $\mu(A \cap B)$ is intended to reflect the uncertainty in equivalence between set elements $f_i \in A$ and $f_j \in B$. We define soft equivalence between elements $f_i \in A$ and $f_j \in B$ as a similarity function or kernel $\mathcal{K}(f_i,f_j)$ ranging from $[0,1]$, where $\mathcal{K}(f_i,f_j)=1$ indicates exact equivalence and $\mathcal{K}(f_i,f_j) = 0$ represents the absence of equivalence. We then define a measure $\mu(A \rightarrow B)$ over a mapping  $A \rightarrow B$ from set $A$ to set $B$ as:
\begin{align}
\mu(A \rightarrow B) 
&= \sum_i^{|A|}\max_{f_j \in B}\mathcal{K}(f_i,f_j).
\label{eq:gen2}
\end{align}
The similarity function $\mathcal{K}(f_i,f_j)$ in Equation~\eqref{eq:gen2} may generally be defined according to the specific representation of set elements $f_i$ and $f_j$. The maximum operator $\max_{f_j \in B}\mathcal{K}(f_i,f_j)$ ensures the bound $\mu(A \rightarrow B) \leq \mu(A)$ via a partial injective mapping from A to B, embodying the intuition that each element $f_i \in A$ has at most one counterpart $f_j \in B$. Note also that max operator may be computed via computationally efficient NN indexing methods. The soft intersection cardinality $\mu(A \cap B)$ may be defined as a symmetrized version of $\mu(A \rightarrow B)$:
\begin{align}
\mu(A \cap B) &= \min \left\{~\mu(A \rightarrow  B),~ \mu(B \rightarrow  A)~\right\}, \notag \\
&= \min \left\{ \sum_i^{|A|}\max_{f_j \in B}\mathcal{K}(f_i,f_j), \sum_j^{|B|}\max_{f_i \in A}\mathcal{K}(f_j,f_i) \right\}.
\label{eq:gen3}
\end{align}
Note that the Jaccard distance measure derived using the cardinality operator $\mu(A \cap B)$ as defined by Equations~\eqref{eq:gen2} and~\eqref{eq:gen3} is symmetric and follows the identity of indiscernibles, and thus may be considered to be a semi-metric. The triangle inequality holds in the case of binary equivalence~\cite{Levandowsky1971DistanceSets} but not generally for soft equivalence relationships. Equation~\eqref{eq:gen3} may be used to define the cardinality of a single set $A$ as $\mu(A)=\mu(A \cap A)=|A|$ both for hard and soft equivalence since, since for a given element $f_i \in A$, $\max_{f_j \in A}\mathcal{K}(f_i,f_j) = \mathcal{K}(f_i,f_i) = 1$. The standard Jaccard index in Equation~\eqref{eq:jaccard} is thus a special case of the soft Jaccard index in the case of a binary-valued kernel $\mathcal{K}(f_i,f_j) \in \{0,1\}$.\\

\noindent {\bf Defining $\boldsymbol{\mathcal{K}(f_i,f_j)}$:} The soft set intersection cardinality $\mu(A \cap B)$ may be adapted to a specific task by defining $\mathcal{K}(f_{i},f_{j})$ according to the representation of elements $f_{i}$. In the context of this article, a set element $f_{i}$ is a 3D scale-invariant keypoint $f_{i} = \{ \bar{a_{i}}, \bar{g_{i}} \}$ as described in~\cite{Toews2013a}, where $\bar{g_{i}}$ and $\bar{a_{i}}$ are descriptors of local keypoint geometry and appearance, respectively. Keypoint geometry $\bar{g_{i}} = \{ \bar{x_{i}},\sigma_{i} \}$ consists of 3D location $\bar{x_{i}}$ and scale $\sigma_{i}$, and appearance $\bar{a_{i}}$ is a vector of local image information, i.e. a rank-ordered histogram of oriented gradients (HOG)~\cite{Toews2009}.\\ 

%In the limiting case of a kernel with binary support $\mathcal{K}(f_i,f_j) \in \{0,1\}$ such as the Iverson bracket $\mathcal{K}(f_i,f_j)=[f_i=f_j]$, the expression in Equation~\eqref{eq:soft_jaccard} is equivalent to the standard Jaccard with hard set equivalence. 

The kernel $\mathcal{K}(f_{i},f_{j})$ operates on keypoint elements $f_{i} = \{ \bar{a_{i}}, \bar{g_{i}} \} \in A$ and $f_{j} = \{ \bar{a_{j}}, \bar{g_{j}} \} \in B$. Here we relax the assumption of hard equivalence using squared exponential kernels with non-zero support, factored into kernels operating separately on local keypoint appearance $\mathcal{K}(\bar{a_{i}},\bar{a_{j}})$ and geometry $\mathcal{K}(\bar{g_{i}},\bar{g_{j}})$ variables:
\begin{align}
    \mathcal{K}(f_{i},f_{j}) &= \mathcal{K}(\bar{a_{i}},\bar{a_{j}})\mathcal{K}(\bar{g_{i}},\bar{g_{j}})
    \label{eq:product_weights}
\end{align}
The factorization in Equation~\eqref{eq:product_weights} is due to the use of descriptors $\bar{a_{i}}$ that are invariant to 7-parameter similarity transforms of the 3D image coordinate system from which geometry $\bar{g_{i}}$ is derived. The two kernels in Equation~\eqref{eq:product_weights} are defined as squared exponential functions as follows. 

The appearance kernel $\mathcal{K}(\bar{a_{i}},\bar{a_{j}})$ is defined by the squared Euclidean distance $\lVert \bar{a_{i}} - \bar{a_{j}} \rVert^{2}_{2}$ between appearance vectors $\bar{a_{i}}$ and $ \bar{a_{j}}$:
\begin{align}
	\mathcal{K}(\bar{a_{i}},\bar{a_{j}}) &= \exp \left( -\frac{\lVert \bar{a_{i}} - \bar{a_{j}} \rVert^{2}_{2}}{\alpha^{2}} \right) \label{eq:appearance_weight}
\end{align}
where in Equation~\eqref{eq:appearance_weight}, $\alpha$ is a bandwidth parameter that may be estimated adaptively as
\begin{equation}
	\alpha = \min_{f_{j} \in \Omega}\lVert \bar{a_{i}} - \bar{a_{j}} \rVert^{2}_{2},~s.t.~\lVert \bar{a_{i}} - \bar{a_{j}} \rVert^{2}_{2} > 0,
\end{equation}
the minimum Euclidean distance between appearance descriptor $\bar{a_{i}} \in A$ and the nearest descriptor $\bar{a_{j}} \in \Omega \setminus A$ within the entire available dataset $\Omega$ excluding $A$. Note this choice of estimator is not strictly symmetric, however it allows the kernel to adapt to arbitrary dataset sizes, shrinking the resolution of prediction as the number of data grows large, and does not affect the symmetry of Equation~\ref{eq:gen3}.

\begin{figure}[!h]
  \centering
  \includegraphics[width=0.45\textwidth,draft=false]{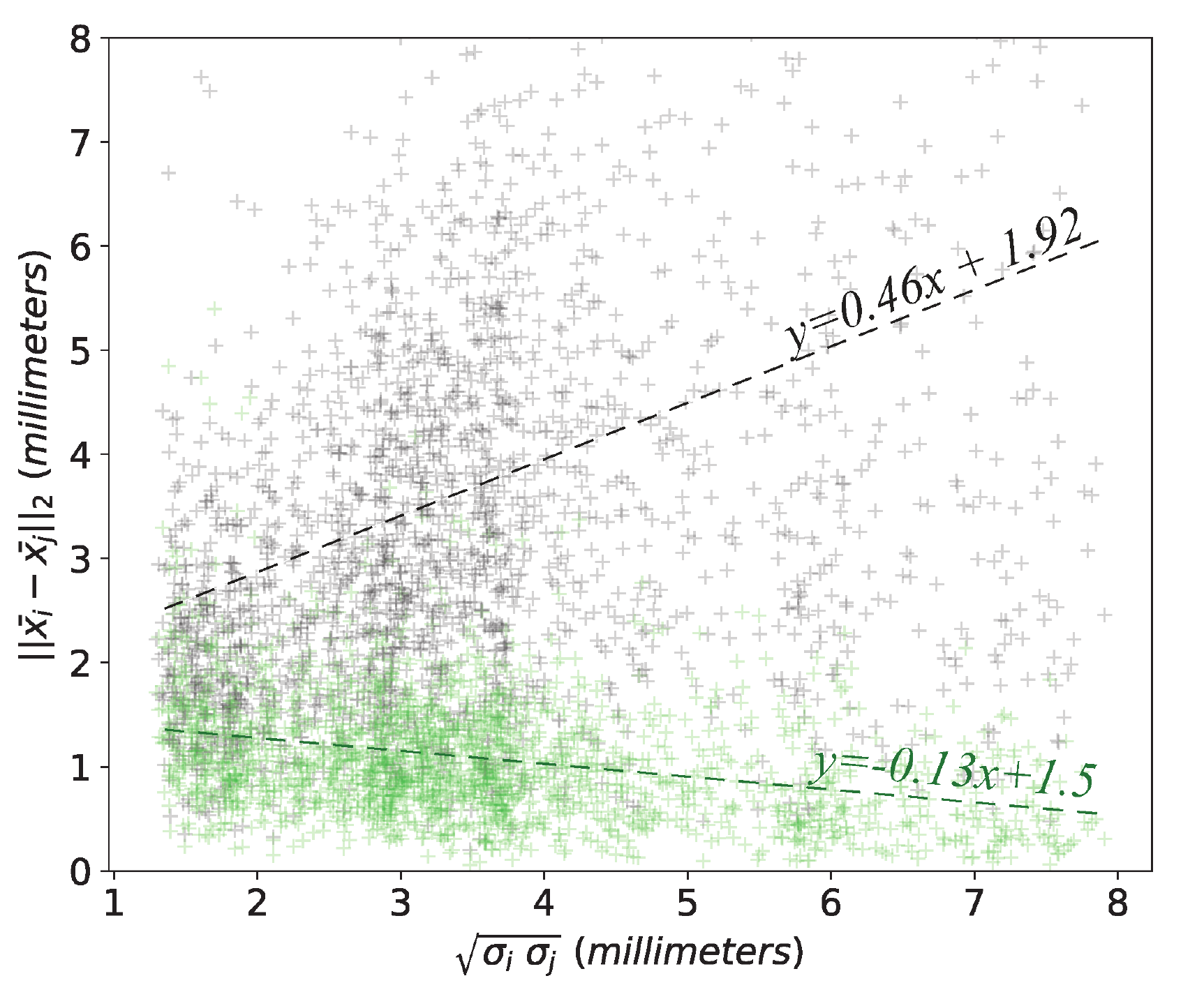}
  \caption{Graph of spatial localization error $\lVert \bar{x_i}-\bar{x_j} \rVert_{2}$ vs. geometric mean scale $\sqrt{\sigma_i~\sigma_j}$. Lines represent the linear regression based on each point set. Note how location error increases with scale (black) due to uncertainty, however this effect is reduced by scale normalization (green) as in Equation~\eqref{eq:geometrical_weight}. This visualization is based on keypoint correspondences between 10 MRI aligned volumes.}
  \label{fig:scale_related_error}
\end{figure}

The geometry kernel $\mathcal{K}(\bar{g_{i}},\bar{g_{j}})$ is novel to this work, and is defined as the product of two kernels, one modeling keypoint location conditional on keypoint scale $\mathcal{K}(\bar{x_{i}},\bar{x_{j}};\sigma_{i},\sigma_{j})$ and the $\mathcal{K}(\sigma_{i},\sigma_{j})$ modeling scale alone  
\begin{align}
    \mathcal{K}(\bar{g_{i}},\bar{g_{j}})=
    \mathcal{K}(\bar{x_{i}},\bar{x_{j}};\sigma_{i},\sigma_{j})\mathcal{K}(\sigma_{i},\sigma_{j})
 \label{eq:geometry_kernel}
\end{align}
These kernels are defined as follows
\begin{align}
    \mathcal{K}(\bar{x_{i}},\bar{x_{j}};\sigma_{i},\sigma_{j}) &= \exp \left( -\frac{\lVert \bar{x_{i}} - \bar{x_{j}} \rVert^{2}_{2}}{\sigma_{i}\sigma_{j}} \right) \label{eq:geometrical_weight} \\
    \mathcal{K}(\sigma_{i},\sigma_{j}) &=  \exp \left( - \log^{2}\left( \frac{\sigma_{i}}{\sigma_{j}} \right) \right) 
    \label{eq:scale_weight}
\end{align}
Kernel $\mathcal{K}(\bar{x_{i}},\bar{x_{j}};\sigma_{i},\sigma_{j})$ in Equation~\eqref{eq:geometrical_weight} penalizes the squared distance between keypoint coordinates within a local reference frame, normalized by a variance proportional to the product of keypoint scales $\sigma_{i}\sigma_{j}$. This variance embodies uncertainty in keypoint location due to scale, and has a computational form that is reminiscent of mass in Newton's law of gravitation or electric charge magnitude in Coulomb's law. Figure~\ref{fig:scale_related_error} demonstrates how this kernel variance normalizes higher localization error associated with keypoints of larger scale. Kernel $\mathcal{K}(\sigma_{i},\sigma_{j})$ in Equation~\eqref{eq:scale_weight} penalizes multiplicative difference between keypoint scales $(\sigma_i,\sigma_j)$.

Note that while the appearance kernel in Equation~\eqref{eq:appearance_weight} is invariant to global similarity transforms due to descriptor invariance, the geometry kernel in Equation~\eqref{eq:geometry_kernel} measures zero-mean keypoint displacement and thus requires data to be aligned within a common spatial reference frame. Alignment may be established via standard subject-to-atlas registration prior to keypoint extraction or from keypoint correspondences after extraction using feature-based alignment~\cite{Toews2013a}. The following workflow is used to compute pairwise distances for a set of $N$ images:
\begin{enumerate}
    \item Extract keypoints from each image
    \item Align keypoints to an atlas template (optional for pre-aligned images) 
    \item Identify k-NN correspondences $(\bar{a_{i}},\bar{a_{j}})$ minimizing the Euclidean distance between descriptors $\| \bar{a_{i}}-\bar{a_{j}} \|$
    \item Compute pairwise image distances by evaluating kernels from k-NN correspondences
\end{enumerate}

\begin{figure}[!h]
  \centering
  \includegraphics[width=0.5\textwidth,draft=false]{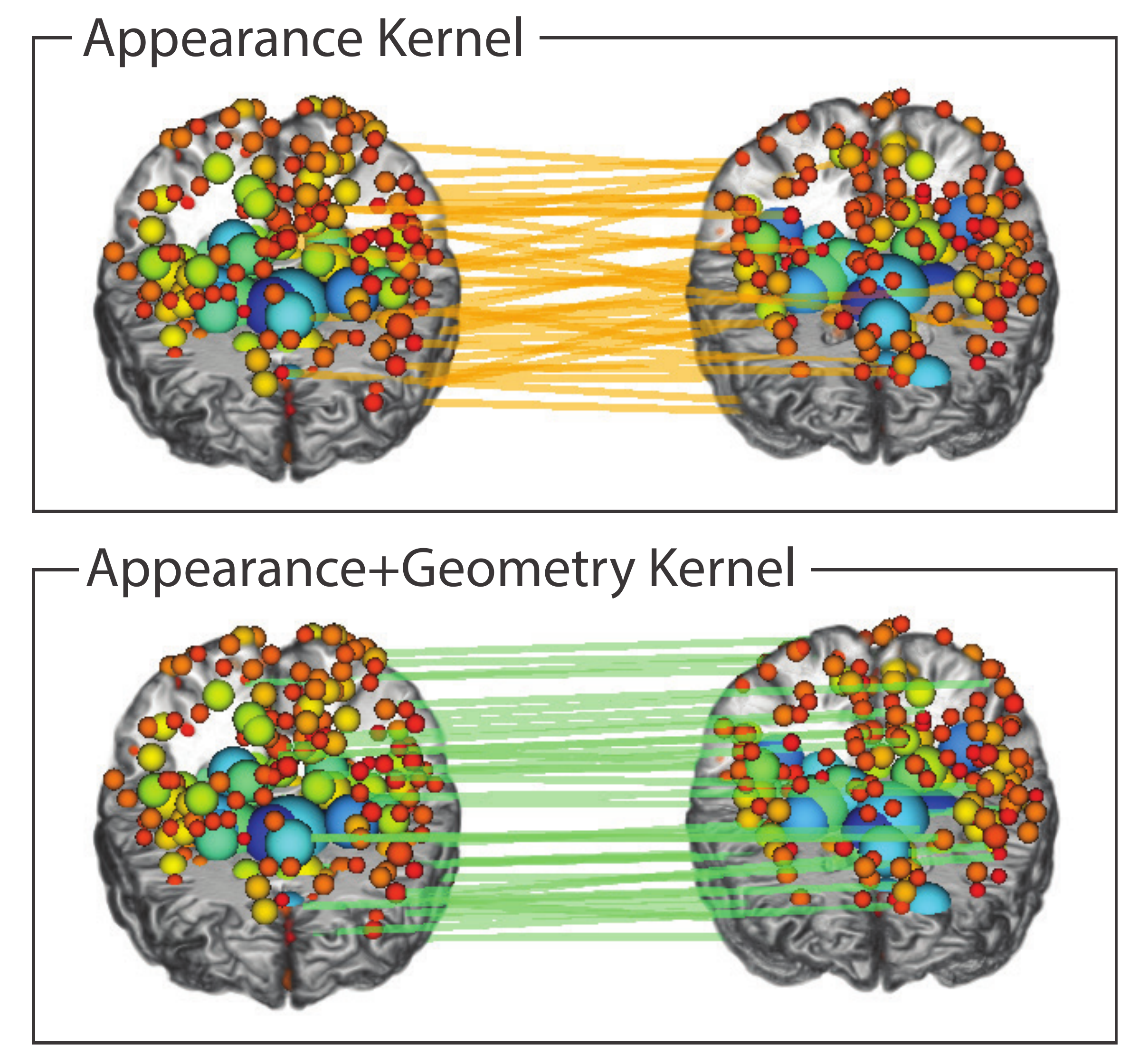}
  \caption{Visualizing the keypoint correspondences driving classification between a pair of identical twin brains, showing the 100 correspondences with the highest SSE kernel values. The appearance kernel alone (upper, Equation~\eqref{eq:appearance_weight}) may include geometrically inconsistent matches (diagonal lines), however these are down weighted when combined with the geometry kernel (lower, Equations~\eqref{eq:geometrical_weight} and~\eqref{eq:scale_weight}), leading to geometrically consistent matches (horizontal lines) and improved prediction performance.}
  %Step I consists of the SIFT feature extraction for each image (via GPU or CPU implementation). In Step II and III, lines represents the 100 most similar features based respectively on kernels defined in equation~\eqref{eq:appearance_weight}, and the combination of kernels from equation~\eqref{eq:appearance_weight},~\eqref{eq:geometrical_weight} and~\eqref{eq:scale_weight}. Step II is achieved in $O(log N)$ complexity  via a k-Nearest Neighbor algorithm. It should be noted that the constraint introduced by our newly defined kernels in equations~\eqref{eq:geometrical_weight} and~\eqref{eq:scale_weight} in Step III results in a more accurate feature matching.}
  \label{fig:workflow}
\end{figure}

Figure~\ref{fig:workflow} shows the effect of our proposed geometry kernel in regularizing and favoring geometrically plausible correspondences for an example twin sibling pair.

\section{Experiments}

Experiments investigate the ability of the Jaccard distance to predict pairwise relationships between whole-brain MRI scans, where relationships include 1) close genetic links between siblings sharing 25-100\% of their polymorphic genes and 2) broad genetic links between non-siblings sharing nominal genetic information due to sex and common racial ancestry. We hypothesize that closer genetic proximity will be reflected in higher pairwise similarity and thus lower Jaccard distance. We expect that pairwise distance based on our proposed soft set equivalence (SSE) $d_{J}(A,B) = -\log J_{SSE}(A,B)$ as defined in Equations~\eqref{eq:soft_jaccard} and~\eqref{eq:gen3} and our geometry kernel as defined by Equations~\eqref{eq:geometrical_weight} and~\eqref{eq:scale_weight} will lead to improved identification of pairwise relationship labels.

Prediction is based on pairwise Jaccard distance and a leave-one-out protocol, analogous to querying a subject MRI in a hospital PACS. Note that this is a deterministic procedure with no explicit training stage, where the primary hyperparameter is the number of nearest neighbors $k$ per keypoint. Our computational workflow, as previously described, follows three steps 1) 3D SIFT-Rank keypoints are extracted from individual pre-processed images, 2) kNN correspondences are identified between keypoints from all $N$ images, and 3) Jaccard distances are evaluated between all image pairs from kernels and kNN correspondences.

\subsection{Data and Computational Details}
Our data set consists of MRI scans of $N=1010$ unique subjects from the Human Connectome Project (HCP) Q4 release~\cite{VanEssen2012}, aged 22-36 years (mean 29 years), acquired from a diverse population including 468 males and 542 females, and 434 unique families. The MRI data are provided as T1w volumes at isotropic 0.7mm voxel resolution and preprocessed via a standard neuroimaging pipeline, including rigid subject-to-atlas registration and skull-stripping. Keypoint extraction requires approximately 3~sec.~/~per image and results in an average of 1,400 keypoints per image for a total of 1,488,065 keypoints. We use a GPU implementation~\cite{pepin2020large} of the 3D SIFT-Rank keypoint algorithm~\cite{Toews2013a} that produces identical keypoints at a 7$\times$ speedup. Approximate kNN correspondences between appearance descriptors are identified across the entire database using efficient KD-tree indexing~\cite{Muja2014}, where lookup requires 0.8~sec.~/~subject for k=200 nearest neighbors on an i7-5600@2.60Ghz machine with 16 GB RAM (1.64 GB used).

Given N=1010 images, there are a total of $N(N-1)/2=509545$ pairwise relationships to be evaluated. Each subject pair is assigned by one of five possible relationship labels $L = \{MZ, DZ, FS, HS, UR\}$ for monozygotic twins (MZ), dizygotic twins (DZ), full non-twin siblings (FS), half-siblings (HS) and unrelated non-siblings (UR), for totals of $\{134, 71, 607, 44, 508689\}$ relationships per label. Family relationship labels are based on mother and father identity and zygoticy (for twins), confirmed via genome-wide single nucleotide polymorphism (SNP) genotyping~\cite{wu20171200}. Note the sparse structure of family relationship labels, where each sibling label (MZ, DZ, FS, HS) is unique to one of 434 families and the vast majority of pairs are UR ($0.998=508689/509545$). The nominal probability of randomly guessing the correct family is approximately $0.0023=1/434$, and the challenge is thus to efficiently and accurately identify the small number of family relationships. Note that evaluating the similarity of $N(N-1)/2$ pairs via brute force image matching or registration becomes computationally intractable for large $N$.

Image alignment is fundamental to the functioning of our proposed geometry kernel, which measures zero-mean deviations in keypoint location within a common spatial reference frame. We thus evaluated prediction following two different subject-to-atlas registration methods: the original 3D rigid transform provided by the HCP mapping images to the standard MNI atlas using the FLIRT algorithm~\cite{Glasser2013TheProject,Jenkinson2002ImprovedImages}, and feature-based alignment (FBA)~\cite{Toews2013a} using keypoint correspondences to estimate a 3D similarity transform mapping images to a subject atlas, for 20 different randomly selected atlases. All alignment solutions were very close to the reference MNI alignment (rotation differed by $4.31^{\circ}\pm 2.84$, translation by $3 mm \pm 3.5$), reflecting a degree of inter-subject variability given different atlases. Most notably, however, was that AUC values were virtually identical for all prediction trials, with a maximum standard deviation of $\epsilon = 0.0017$ (see Table~\ref{tab:hard_soft_ROC}), indicating that prediction was not sensitive to the alignment solution used. The results we report are thus consistent with the reference MNI alignment provided with the HCP data.

\subsection{Close genetic proximity: Siblings}

\begin{figure*}[!h]
  \centering
  \includegraphics[width=0.85\textwidth,draft=false]{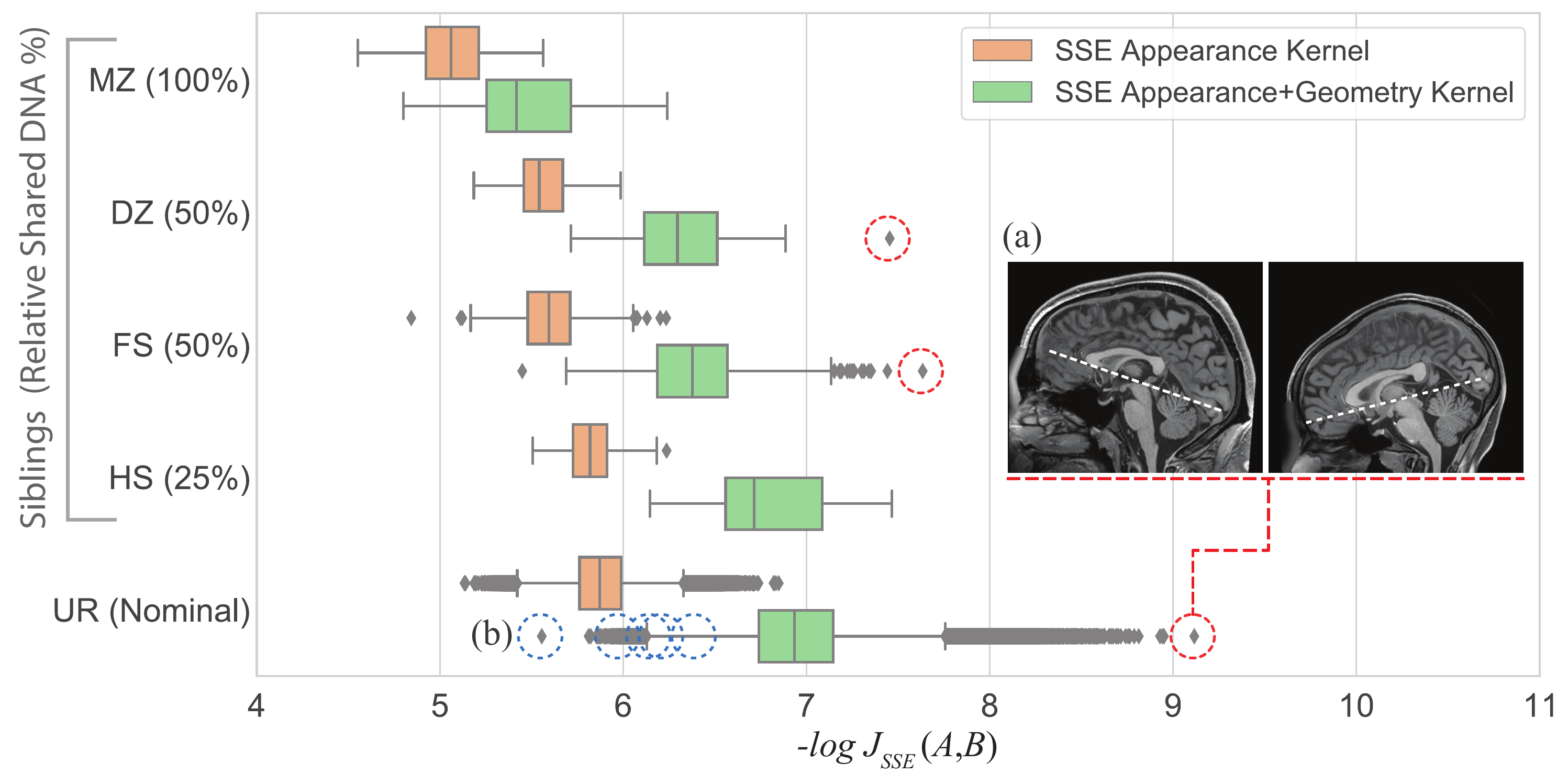}
  \caption{Distributions of Jaccard distances $-\log J_{SSE}(A,B)$ conditioned on pairwise labels $L = \{MZ, DZ, FS, HS, UR\}$, comparing kernels with and without geometry (green vs. orange). Note that the distance generally increases with genetic separation, and that geometry kernels increase separation between sibling relationship labels. Outliers, i.e. pairs outside of their expected relationship distance distributions, can be easily identified and inspected for irregularity, e.g. (a) unusually high distances may indicate spatial misalignment (red circles) and (b) unusually low UR distances may indicate cases of incorrect family labels (blue circles).
  %like the blue circles, indicating pairs of self-reported family members flagged as unrelated due uncertain genotyping~\cite{wu20171200}.
  } 
  \label{fig:rel_distances}
\end{figure*}

Figure~\ref{fig:rel_distances} shows distributions of pairwise Jaccard distance conditioned on pairwise relationships. Our proposed kernel based on appearance and geometry (Figure~\ref{fig:rel_distances}, green) generally increases separation between different sibling relationships in comparison to appearance only (Figure~\ref{fig:rel_distances}, orange) as in~\cite{Chauvin2020NeuroimageRelatives}. A two-tailed Kolmogorov-Smirnoff (KS) test shows all distributions to be significantly different ($p-value < 1e-10$) except those of DZ and FS siblings sharing approximately 50\% of their genes ($p-value = 0.0199$). Furthermore, pairwise KS statistics were all lower for appearance+geometry kernels compared to appearance only, indicating that the geometry kernels lead to increased separation of relationship categories. The Jaccard distance may be rapidly evaluated, and pairs which are outliers from their expected distributions may be easily flagged and inspected for irregularities. For example, the highest Jaccard distance outliers indicated noticeable spatial misalignment between a small number of subject pairs (e.g. Figure~\ref{fig:rel_distances}a. The lowest distance outliers for UR pairs may indicate potential sibling relationships. The blue circles near Figure~\ref{fig:rel_distances}b are pairs involving 3 subjects flagged as UR due to genotyping errors (including a MZ twin)~\cite{wu20171200}. Here, nearest neighbor Jaccard distance was used to correctly predict the self-reported families for all three cases, a result confirmed by the HCP.

\begin{figure*}
  \centering
  \includegraphics[width=0.8\textwidth,draft=false]{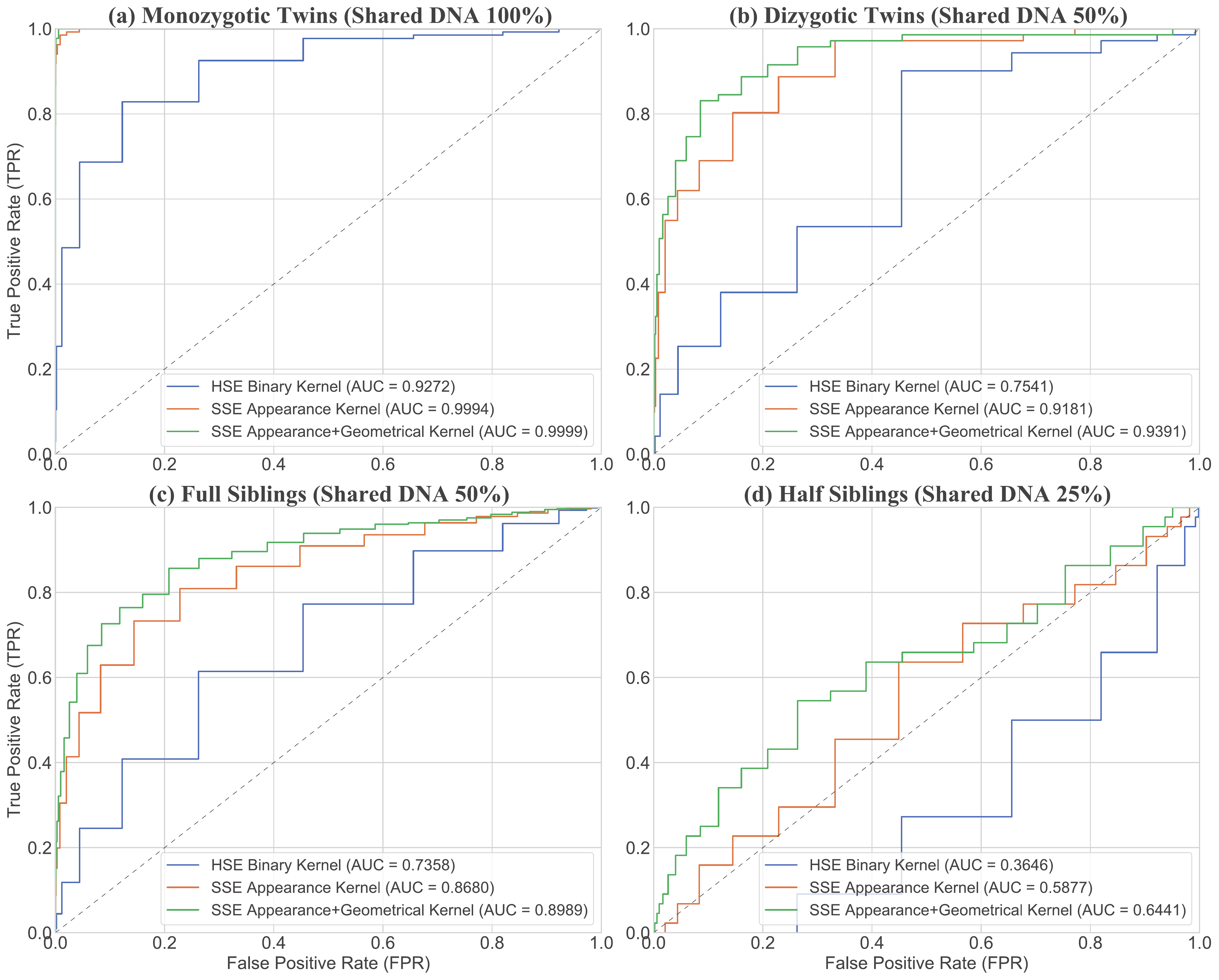}
  \caption{ROC curves for sibling identification based on Jaccard distance for (a) MZ, (b) DZ, (c) FS and (d) HS pairs, comparing three kernels for k=200 NNs. Note that the prediction AUC is always highest for our proposed kernel combining appearance and geometry (green), and decreases as expected with the degree genetic and developmental difference, e.g. in the order MZ, DZ, FS, HS.}
  \label{fig:roc_curves}
\end{figure*}

As sibling pairs exhibit significantly lower distance than UR pairs, we investigate the degree to which they can be distinguished from unrelated pairs based on a simple distance threshold. Figure~\ref{fig:roc_curves} shows the Receiver Operating Characteristic (ROC) curves for sibling relationships based on distance, comparing our SSE kernels for appearance and geometry, appearance only and the binary HSE kernel. Table~\ref{tab:hard_soft_ROC} quantifies the improvement of SSE vs HSE, for various numbers of keypoint nearest neighbors (20, 100, 200), where the highest area-under-curve (AUC) values are obtained for SSE with k=200. The performance of SSE kernels generally increases with the number of NNs, and is always superior for combined appearance and geometry kernels. Conversely, HSE classification performance decreases, even falling below the diagonal in the most challenging case of HS pairs in Figure~\ref{fig:roc_curves}d, as additional hard correspondences appear to accumulate systematically on a subset of typical but unrelated subjects. For completeness, the highest UR prediction result was AUC$=0.951$ for SSE, appearance and geometry kernels, k=200.
 
\begin{table*}
\begin{center}
\bgroup
\def\arraystretch{1.5}
\begin{tabular}{|cc|ccc|ccc|ccc}
    \hline
    \multicolumn{2}{|P{2cm}|}{\multirow{2}{*}{\backslashbox{Label}{Kernel}}}&\multicolumn{3}{|c|}{\textit{k = 20}}&\multicolumn{3}{|c|}{\textit{k = 100}}&\multicolumn{3}{|c|}{\textit{k = 200}}\\ 
    \cline{3-11}
    \multicolumn{2}{|P{2cm}|}{}&\multicolumn{1}{|P{1.2cm}|}{\textit{HSE}~\cite{Toews2016}}&\multicolumn{1}{|P{1.2cm}|}{\textit{SSE\textsubscript{App.}}\cite{Chauvin2020NeuroimageRelatives}}&\multicolumn{1}{P{1.5cm}|}{\textit{SSE\textsubscript{App. + Geo.}}}&\multicolumn{1}{|P{1.2cm}|}{\textit{HSE}~\cite{Toews2016}}&\multicolumn{1}{|P{1.2cm}|}{\textit{SSE\textsubscript{App.}}\cite{Chauvin2020NeuroimageRelatives}}&\multicolumn{1}{P{1.5cm}|}{\textit{SSE\textsubscript{App. + Geo.}}}&\multicolumn{1}{|P{1.2cm}|}{\textit{HSE}~\cite{Toews2016}}&\multicolumn{1}{|P{1.2cm}|}{\textit{SSE\textsubscript{App.}}\cite{Chauvin2020NeuroimageRelatives}}&\multicolumn{1}{P{1.5cm}|}{\textit{SSE\textsubscript{App. + Geo.}}}\\
    \thickhline
    \multicolumn{2}{|P{2cm}|}{\textbf{Monozygotic}}&\multicolumn{1}{|P{1.2cm}|}{0.9983}&\multicolumn{1}{P{1.2cm}|}{0.9993}&\multicolumn{1}{P{1.5cm}|}{\textbf{0.9996 $\bm{\pm~\epsilon}$}}&\multicolumn{1}{|P{1.2cm}|}{0.9544}&\multicolumn{1}{P{1.2cm}|}{0.9995}&\multicolumn{1}{P{1.5cm}|}{\textbf{0.9998 $\bm{\pm~\epsilon}$}}&\multicolumn{1}{|P{1.2cm}|}{0.9272}&\multicolumn{1}{P{1.2cm}|}{0.9994}&\multicolumn{1}{P{1.5cm}|}{\textbf{0.9999 $\bm{\pm~\epsilon}$}}\\
    \hline
    \multicolumn{2}{|P{2cm}|}{\textbf{Dizygotic}}&\multicolumn{1}{|P{1.2cm}|}{0.8922}&\multicolumn{1}{P{1.2cm}|}{0.8825}&\multicolumn{1}{P{1.5cm}|}{\textbf{0.9044 $\bm{\pm~\epsilon}$}}&\multicolumn{1}{|P{1.2cm}|}{0.8018}&\multicolumn{1}{P{1.2cm}|}{0.9034}&\multicolumn{1}{P{1.5cm}|}{\textbf{0.9250 $\bm{\pm~\epsilon}$}}&\multicolumn{1}{|P{1.2cm}|}{0.7541}&\multicolumn{1}{P{1.2cm}|}{0.9181}&\multicolumn{1}{P{1.5cm}|}{\textbf{0.9391 $\bm{\pm~\epsilon}$}}\\
    \hline
    \multicolumn{2}{|P{2cm}|}{\textbf{Full-Sibling}}&\multicolumn{1}{|P{1.2cm}|}{0.8423}&\multicolumn{1}{P{1.2cm}|}{0.8433}&\multicolumn{1}{P{1.5cm}|}{\textbf{0.8753 $\bm{\pm~\epsilon}$}}&\multicolumn{1}{|P{1.2cm}|}{0.7611}&\multicolumn{1}{P{1.2cm}|}{0.8569}&\multicolumn{1}{P{1.5cm}|}{\textbf{0.8888 $\bm{\pm~\epsilon}$}}&\multicolumn{1}{|P{1.2cm}|}{0.7358}&\multicolumn{1}{P{1.2cm}|}{0.8680}&\multicolumn{1}{P{1.5cm}|}{\textbf{0.8989 $\bm{\pm~\epsilon}$}}\\
    \hline
\end{tabular}
\egroup
\end{center}
\caption{Area Under the Curve (AUC) values for sibling classification, comparing kernels and numbers of nearest neighbors (k). $\epsilon = 0.0017$ is the maximum AUC standard deviation observed from 21 different subject-to-atlas registration trials. Note $SSE_{App+Geo}$ kernels significantly outperform $SSE_{App}$ and $HSE$ for the more difficult DZ and FS pairs.}
\label{tab:hard_soft_ROC}
\end{table*}

\subsection{Distant genetic proximity: unrelated subjects}

\begin{figure*}[!h]
  \centering
  \includegraphics[width=0.85\textwidth,draft=false]{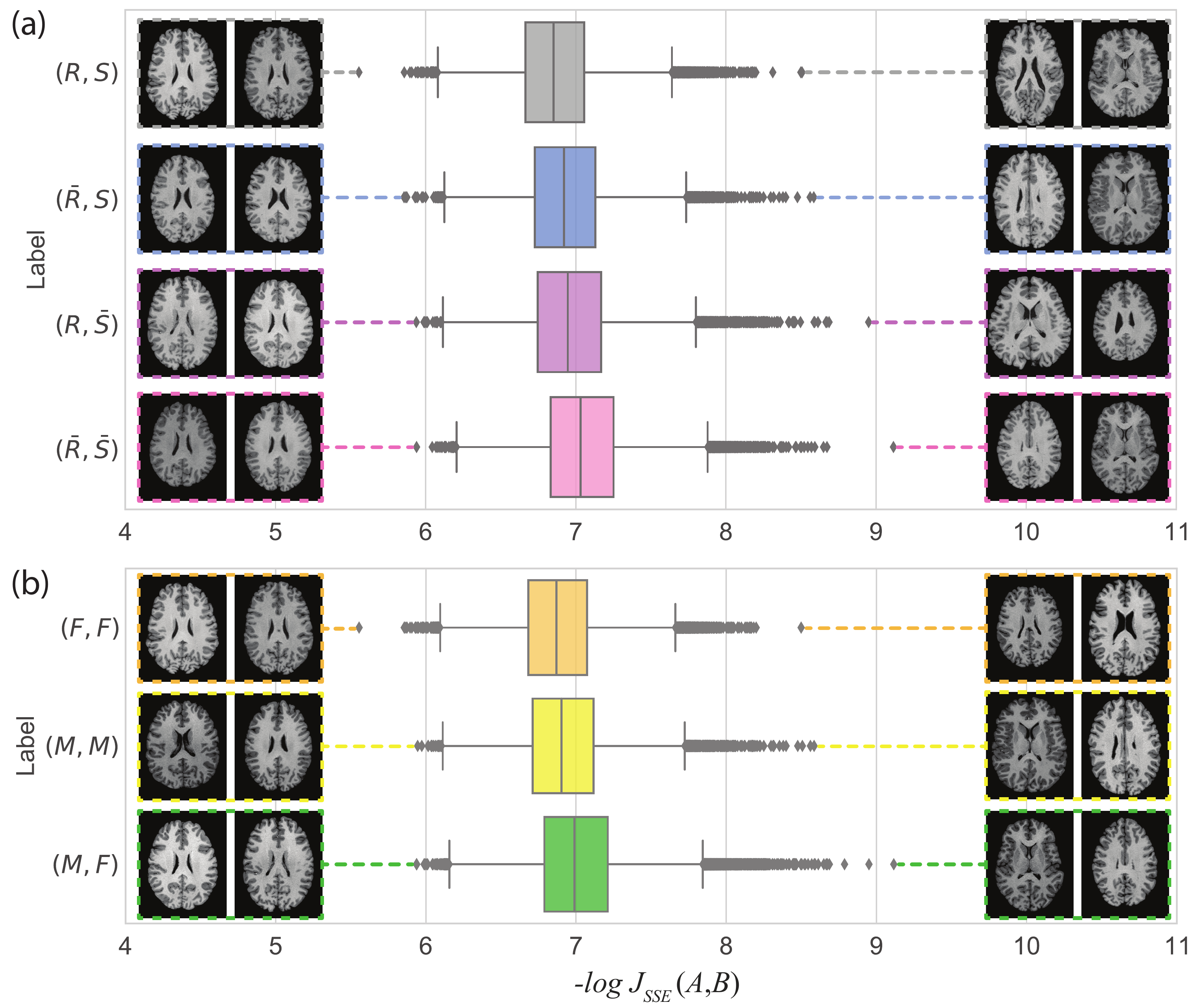}
  \caption{Jaccard distance $-\log J_{SSE}(A,B)$ distributions between unrelated pairs conditioned on shared demographic information. \textbf{(a)} same race, same sex $(R,S)$; different race, same sex $(\bar{R},S)$; same race, different sex $(R,\bar{S})$, and different race, different sex $(\bar{R},\bar{S})$ and \textbf{(b)} Female-Female $(F,F)$, Male-Male $(M,M)$, Male-Female $(M,F)$. Image pairs corresponding to minimum and maximum distances for each distribution are shown for visualization.}
  \label{fig:gender_race}
\end{figure*}

Unrelated (UR) subject pairs share nominal amounts of genetic information, with subtle similarities due to demographic factors such as common sex and ancestral race. We thus expect whole-brain distance to be lowest for pairs of the same race and sex (R,S), highest for different race and sex ($\overline{\mbox{R}}$,$\overline{\mbox{S}}$), and intermediate for either same race (R,$\overline{\mbox{S}}$) or sex ($\overline{\mbox{R}}$,S). While the mean distances for conditional distributions in Figure~\ref{fig:gender_race} generally increase with differences in demographic labels, there are many exceptions where pairs with different labels exhibit lower distance than those with the same labels. We note this is consistent with pairwise genetic differences, where pairs of unrelated individuals from different populations may often exhibit higher genetic similarity than those from the same population~\cite{witherspoon2007genetic}.

\begin{figure}[!h]
  \centering
  \includegraphics[width=0.5\textwidth,draft=false]{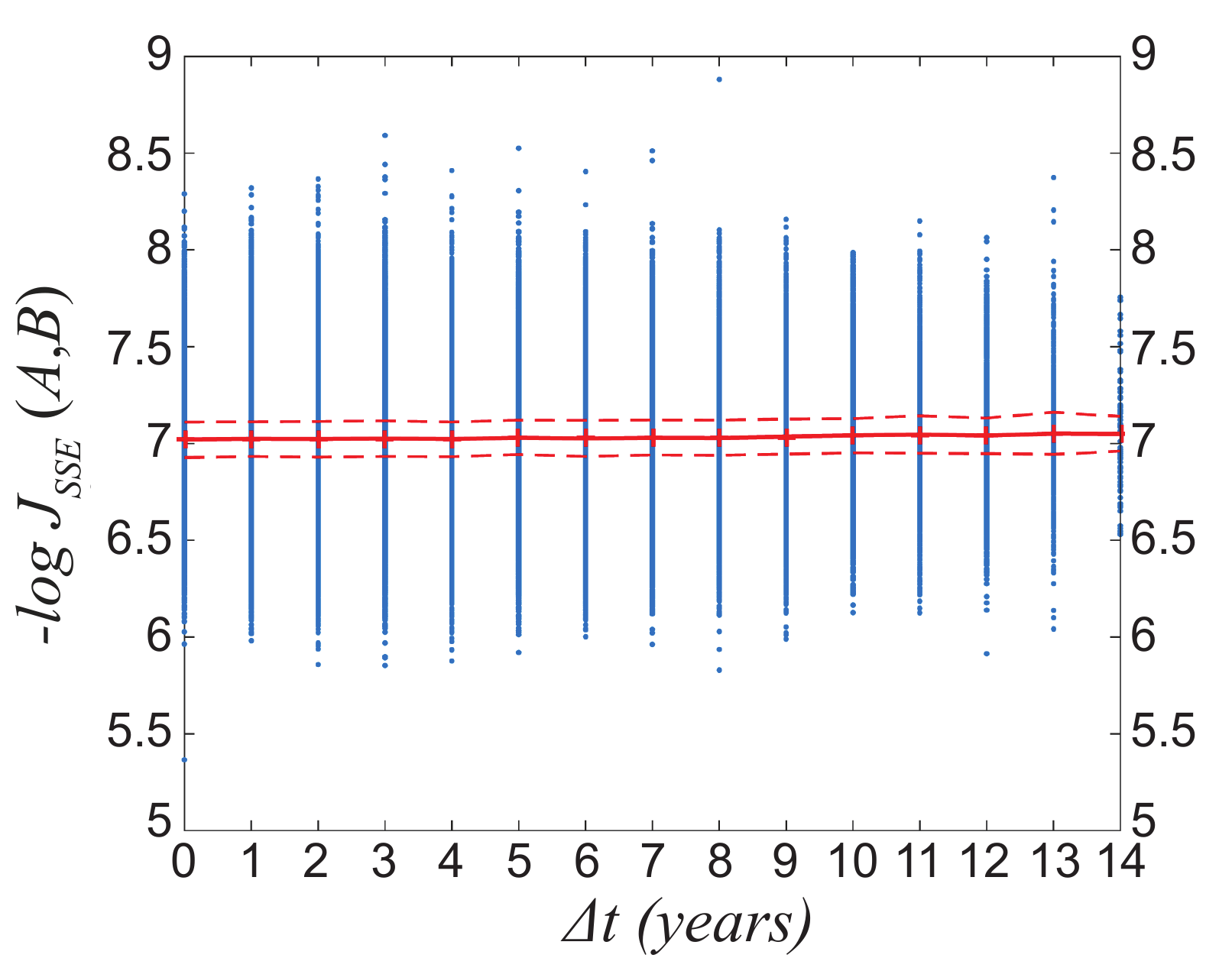}
  \caption{Distributions of Jaccard distance $-\log J_{SSE}(A,B)$ conditioned on age difference $\Delta t$ between unrelated pairs $(A,B)$ conditioned on age difference $\Delta t$. The mean (solid red line) and standard deviation (dashed red lines) are plotted for each $\Delta t$. Age difference has no significant impact on the distance for unrelated healthy young adult brains (age 22-36 years), eliminating a possible confounding factor.}
  \label{fig:age_difference}
\end{figure}

Age difference between subjects is a potential confounding factor in whole brain Jaccard distance, and we plotted the variation of distance vs. age difference in Figure~\ref{fig:age_difference}. Distance distributions are virtually identical across the HCP subject age range spanning 22-36 years of age, indicating that age difference is not a major confound in this relatively young and healthy HCP cohort where brain morphology is relatively stable. Note that the Jaccard distance was found to increase with age difference in older subjects due to natural aging and neurodegenerative disease in~\cite{Chauvin2020NeuroimageRelatives}.
%similarly in rapid neurodevelopment over the infant range in~\cite{Toews2012Feature-basedMRI}.  

\subsection{Group Prediction: Sex}

While the pairwise Jaccard distance is highly informative regarding sibling relationships, it is insufficient for predicting group labels such as sex, age or disease. A simple modification can be used to predict group labels, by evaluating the distance between a single keypoint set A and supersets formed by the union of group members (excluding subject A), e.g. keypoints for {\em all Males} $d_{J}(A,\mathcal{M})$ and {\em all Females} $d_{J}(A,\mathcal{F})$. Supersets $\mathcal{M}$ and $\mathcal{F}$ are solely composed of unrelated subjects to avoid potential biases due to family relationships. These distances are combined in a basic linear classifier with a single threshold parameter $\tau$ to adjust for differences in the numbers of keypoints per group based on the following equation:
\begin{align}
   Class(A) = &
   \begin{cases}
    Female& \text{if } d_{J}(A,\mathcal{F}) - d_{J}(A,\mathcal{M}) + \tau > 0\\
    Male& \text{otherwise.}
   \end{cases}
   \label{eq:classification}
\end{align}

\begin{figure}[!h]
  \centering
  \includegraphics[width=0.45\textwidth,draft=false]{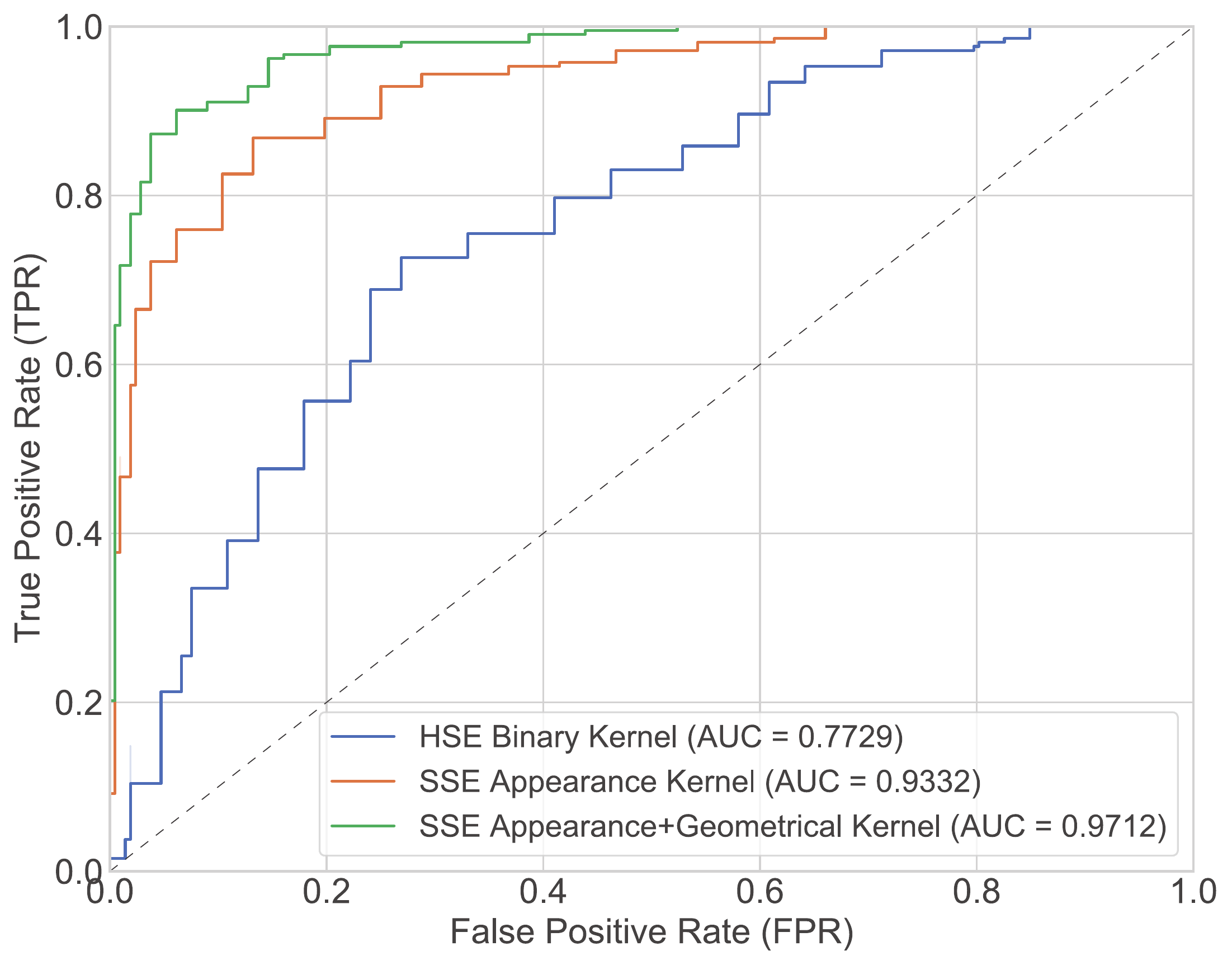}
  \caption{ROC curves and AUC for sex prediction based on individual-to-group Jaccard distance, comparing binary, appearance only, and combined appearance and geometry kernels.}
  \label{fig:ROC_male_female}
\end{figure}

Figure~\ref{fig:ROC_male_female} shows ROCs curves for sex prediction obtained by varying $\tau$ over the range $(-\infty,\infty)$, comparing Jaccard distance computed with binary, appearance only and combined appearance and geometry kernels. A possible confound is brain size, which is on average slightly larger for males than females~\cite{Eliot2021DumpSize}. As the appearance kernel is invariant to image scale, the AUC=0.93 reflects prediction accuracy independently of image size. Our proposed combined approach achieves the highest AUC=0.97, again outperforming other options.

\section{Discussion}

In this paper we propose a novel, highly specific distance measure between volumetric images represented as invariant keypoints. Our measure generalizes the Jaccard index to account for soft equivalence between keypoints, and a novel kernel estimator is proposed to model keypoint geometry in terms of location and scale within normalized image space. The soft Jaccard index is used as a distance measure to predict pairwise relationships between MRI brain scans of 1010 subjects from 434 families, including siblings and twins, and significantly improves upon previous work based on binary equivalence and appearance descriptors~\cite{Chauvin2020NeuroimageRelatives,Kumar2018Multi-modalFramework}. We report the first results for predicting relationships from medical images, a new task, where monozygotic twins can be identified with virtually perfect accuracy. A minor modification allows the Jaccard distance to predict group labels such as sex with high accuracy. Our geometry kernel requires spatial normalization, however trials involving various linear subject-to-atlas alignment solutions including robust keypoint-based alignment~\cite{Toews2013a} show that the prediction performance to be insensitive to the specific alignment solution used. The improved prediction afforded by our geometry kernel thus suggests that family members tend to align in a consistent manner regardless of the solution used.

Our method promises to be a useful tool for curating large medical image datasets for precision medicine and research purposes. A memory-based model using efficient and robust algorithms for 3D keypoint extraction and indexing~\cite{Toews2013a,toews2013feature} allows for fine-grained comparisons between $O(N^2)$ image pairs in $O(N~\log~N)$ computational complexity. The Jaccard distance may thus be used to rapidly validate relationship labels, e.g. unexpectedly low distance may indicate related individuals and unexpectedly high distance may indicate unrelated individuals or spatially misaligned pairs. Previous work identified mislabelled scans of individuals in large neuroimage datasets~\cite{Chauvin2020NeuroimageRelatives}, here our improved method allowed us predict the self-reported families of three subjects labelled as unrelated due to inconclusive genotyping~\cite{wu20171200}, on the basis of nearest neighbor soft Jaccard distance. This exceptional result was confirmed by the Human Connectome Project, and serves as a concrete example as to how our method can be used to validate labels associated with medical imaging data.

The task of pairwise relationship prediction may be viewed as the finest grain of categorization, where the number of unique pairwise relationship labels is linear $O(N)$ in the number of data $N$, e.g. 434 families from N=1010 images, as opposed to typical classification where all data are associated with a small number of labels (e.g. male, female). Pairwise prediction thus represents a challenge for ubiquitous deep neural network methods, which generally require large numbers of training data per category, and training for specific modalities and body parts. In contrast, generic 3D SIFT keypoints may be used as-is with arbitrary imaging modalities and contexts with no training. In future work, generic keypoint correspondences here could potentially be used to train domain-specific keypoint models~\cite{Yi2016LIFT:Transform,Detone2018SuperPoint:Description}. Auto-encoders used for anomaly detection could potentially be adapted to generate subject-specific codes between pairs of images~\cite{Baur2021AutoencodersStudy}. Analysis beyond pairs of subjects could be generalized via graph theory to model the clique structure of families~\cite{blair1993introduction}. All code required to reproduce our results may be obtained at https://github.com/3dsift-rank.

\bibliographystyle{IEEEtran}
\bibliography{IEEEabrv,TMI19,mendeley}

\end{document}